\documentclass{aa}  
\usepackage{graphicx}
\usepackage{txfonts}
\begin{document}
   \title{HD~75289Ab revisited}

   \subtitle{Searching for starlight reflected from a hot Jupiter}

   \author{F. Rodler
          \inst{1,2}
          \and
          M. K\"urster\inst{1}
          \and
          T. Henning\inst{1}
          }

   \offprints{rodler@mpia.de}

   \institute{Max-Planck-Institut f\"ur Astronomie, 
              K\"onigstuhl 17, 69117 Heidelberg, Germany
         \and
             Institut f\"ur Astronomie, Universit\"at Wien,
              T\"urkenschanzstrasse 17, A-1180 Vienna, Austria
             }

   \date{Received ?; accepted ?}

 
  \abstract
  {}
   {We attempt to detect starlight reflected from a hot Jupiter, orbiting the
   main-sequence star HD~75289Ab. We report a revised analysis of observations
   of this planetary system presented previously by another research group.
   }
   {We analyse high-precision, high-resolution spectra, collected over four
   nights using UVES at the VLT/UT2, by way of data
   synthesis. We try to interpret our data using different atmospheric models for hot Jupiters.}
   {We do not find any evidence for reflected light, and, therefore, establish
     revised upper limits to the planet-to-star flux ratio at the
   99.9\% significance level. At high orbital inclinations, where the best
     sensitivity is attained, we can limit the relative reflected
     radiation to be less than $\epsilon =
   6.7\times10^{-5}$ assuming a grey albedo, and
     $\epsilon=8.3\times10^{-5}$ assuming an Class IV function, respectively.
       This
     implies a geometric albedo smaller than $p=0.46$ and $p=0.57$, for the
     grey albedo and the Class IV albedo shape, respectively, assuming a planetary radius of $1.2~\rm{R_{Jup}}$.} 
   {} 


   \maketitle


   \section{Introduction}

   Since the detection of the first exoplanet orbiting a solar-type star, more
   than 270 exoplanets have been detected. So-called hot Jupiters - giant
   planets only a few solar radii away from their host stars - provide the opportunity to detect starlight reflected from these planets. Five
   extended campaigns for the search for reflected light were completed by
   different groups,
   all resulting in non-detections (e.g. Leigh et al.~2003b, and
   references therein). Upper
   limits to the planet-to-star flux ratio and to the geometrical albedo of these
   planets were established, which provided important constraints to 
   models of the planetary atmospheres by Sudarsky et al. (2000, 2003). As a result,
   models that predicted a high reflectivity for the planetary
   atmosphere could be ruled out for some of the studied planets.

   Here, we present a reanalysis of observations of the
   planetary system of HD~75289A, conducted by
   Leigh et al. (2003a), over four nights in January 2003, using the UV-Visual
   Echelle Spectrograph (UVES) mounted 
   on the VLT/UT2 at Cerro Paranal in Chile. These authors attempted to detect starlight reflected from
   the hot Jupiter, but were unable to find evidence for the planetary signal:
   they placed a 99.9\% confidence upper limit on the planet-to-star flux ratio of
   $4.17\times10^{-5}$, for the spectral range $\lambda= 402.33$ to
   $522.13~\rm{nm}$, and for an
   orbital inclination $i=60^{\circ}$. 

   We noticed, however, that this upper
   limit was based on erroneous orbital phase information for the planet. In
   this article, we correct the upper limit to the planet-to-star flux ratio
   determined by Leigh et al. (2003), using using our implementation of
   the modeling approach introduced by Charbonneau et al. (1999). Section~2 describes the
   basic ideas of the search for reflected light. Section~3 provides a brief
   overview of the science data, while Section~4 
   provides a
   detailed description of sophisticated data processing implemented by using the data modeling
   approach. Finally, in Section~5 we present our corrected upper limits to the 
   planet-to-star flux ratio.


  \section{Reflected light}

  \subsection{Photometric Variations}
\begin{figure}
   \centering
   \includegraphics[angle=-90,width=9cm]{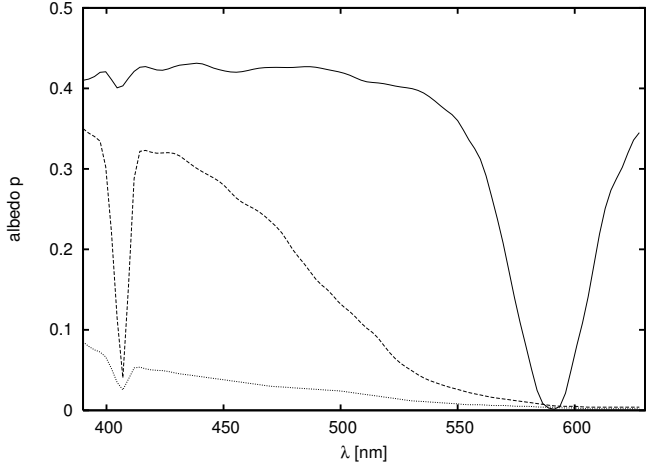}
      \caption{Different albedo spectra of atmospheric models (taken from Sudarsky et
   al. 2000) are shown. The irradiated (dots) and isolated (dashed) Class IV
               models describe atmospheres of planets with temperatures 
               $T_{\rm{eff}}\approx1300~\rm{K}$. In contrast to the isolated model,
               the irradiated model
               assumes that no reflective clouds exist in the upper layer of
               the planetary atmosphere, which results in a very low
               albedo. 
               The solid line depicts a Class V roaster, describing
               the atmosphere of a planet with $T_{\rm{eff}}\ge1500~\rm{K}$, having a high
               reflective silicate cloud deck in the upper layers of the atmosphere.  }
         \label{fig:albedo}
   \end{figure}

  For exoplanets, the enormous brightness contrast between the star and 
  the planet constitutes considerable challenge when attempting to observe some 
  kind of direct signal from the planet.
  For close-in planets such as hot
  Jupiters, the main contribution to the optical flux originates in the
  reflected starlight and not the intrinsic luminosity (Seager et al. 2000).
  High-resolution spectroscopy in the
  optical utilises the fact that the observed spectrum reflected from the 
  planet is essentially a copy of the rich  stellar absorption-line spectrum. 
  Basically, this spectrum is shifted in wavelength according to the orbital
  radial velocity (RV) of the planet and scaled down in 
  brightness by a factor of a few times $10^4$ for hot Jupiters.
 According to Charbonneau et al. (1999), the amount of starlight
  reflected from a planet which is fully illuminated can be described by
\begin{equation} \label{equ:2}
     \epsilon(\lambda) = p(\lambda) \left(\frac{R_{\rm{p}}}{a}\right)^2 \rm{,}
  \end{equation}
  where $p(\lambda)$ denotes the albedo of the planet as a function of the
  wavelength $\lambda$, $R_{\rm{p}}$ the planetary radius and $a$ the
  star-planet separation. Figure \ref{fig:albedo} shows different albedo
  spectra, derived using different planetary-atmosphere models from Sudarsky et
  al. (2000).
The planetary
  radius of HD~75289Ab is unknown; it can, however, be estimated from the
  transiting planets, which provide an exact
  determination of their masses and radii (see Section~2.3). The orbital
  radius can be tightly constrained using Kepler's third law.
 
  In most cases, the planet does not appear to be fully illuminated. Consequently, the
  observed reflected light  is reduced, depending on the model describing the
  scattering behaviour of the atmosphere, its orbital inclination
  $i \in [0^{\circ},90^{\circ}]$ and the orbital phase $\phi \in [0, 1]$ of the planet. 
  We note that we adopt the convention that $\phi=0$ represents inferior
  conjunction  of the planet (for $i=90^{\circ}$, it would be the transit position).  We apply an empirical
  scattering model of the atmospheres of Jupiter and Venus (Hilton 1992),
  which can be approximated by

  \begin{equation} \label{equ:3}
    \mu(\phi, i)= 10^{-0.4 \zeta(\alpha)}  \rm{,}
  \end{equation}
  where
  \begin{equation} \label{equ:5}
    \zeta(\alpha) = 0.09 \left(\frac{1.8~\alpha}{\pi}\right) +  2.39
    \left(\frac{1.8~\alpha}{\pi}\right)^2 -0.65
    \left(\frac{1.8~\alpha}{\pi}\right)^3  
  \end{equation}
  and the phase angle $\alpha$:
  \begin{equation} \label{equ:1}
    \cos \alpha = -\sin i \cos 2\pi\phi\rm{.}
  \end{equation}
%
%
%
  Finally, the flux of the reflected light from the planet at the orbital
  phase $\phi$ follows from equations
  \ref{equ:2} and \ref{equ:3}:
  \begin{equation} \label{equ:4}
    f(\phi, i, \lambda) = \epsilon(\lambda)~\mu(\phi, i) \rm{.}
  \end{equation}
%
%

 \subsection{Doppler shifts}


 The planet orbiting its host star produces not only a flux variation 
 (Equation \ref{equ:4}), but also a Doppler shift of the
 stellar spectrum reflected from the planet. The RV semi-amplitude $K_{\rm p}$ of that shift
 depends on the orbital inclination $i$, which is unknown for non-transiting
 planets. $K_{\rm{p}}$ can be expressed by
  \begin{equation} \label{equ:doppler}
    K_{\rm{p}} = K_{\rm{s}} \frac{M_{\rm{s}}}{M_{\rm{p}}~\sin i} \sin i
    \rm{,}
  \end{equation}
  where $K_{\rm{s}}$ is the RV semi-amplitude of the star, and
  $M_{\rm{s}}$ and $M_{\rm{p}} \sin i$ are the stellar mass and the minimum
  mass of the planet,
  respectively. The largest possible amplitude $K_{\rm{p,max}}$ occurs at the
  orbital inclination $i=90^{\circ}$.

  The instantaneous RV shift of the planetary signal with respect to the star depends 
  on the orbital phase
  $\phi$, 
  \begin{equation} \label{equ:doppler1}
    V_{\rm{p}} = K_{\rm{p}} \sin 2\pi\phi
    \rm{.}
  \end{equation}

  \subsection{HD~75289A and its planet} \label{planet}
  
  High-precision RV measurements revealed the existence of a hot
  Jupiter orbiting the G0-type main-sequence star
  HD~75289A (Udry et al. 2000). 
Table~\ref{tab:hd75289} summarises the
  parameters of the planet and its host star. 
  We note in passing that the system also contains a faint low-mass stellar
  component, separated by $\approx 621~\rm{AU}$ from the primary (Mugrauer et
  al. 2004).

  \begin{table}
    \caption{Parameters of the star HD~75289A and its planetary system}             
    \label{tab:hd75289}      
    \begin{center}                         
    \begin{tabular}{l r l l }        
      \hline\hline                 
      Parameter & Value & Error &Ref. \\
      \hline                        
      Star: \\
      Spectral type & G0 V  & & G89 \\ 
      $V~(mag)$        & 6.35 & & VF05\\
      $d~ (\rm{pc})$ & 28.94 & 0.47 & VF05\\
      $M_{\rm{s}}~ (\rm{M_{\odot}})$ & 1.23 & 0.10 & VF05 \\
      $R_{\rm{s}}~(\rm{R_{\odot}})$ & 1.249 & 0.022 & VF05\\
      $P_{\rm{rot}}~({\rm d})$ & 16.0 & 3.0 & U00 \\
      $v~ \sin i~ (\rm{km~s^{-1}})$ & 4.14 & & B06 \\
      Age (Gyr) & $2.2$  & & VF05 \\
      \hline
      Planet:\\
      $M_{\rm{p}} \sin i~(\rm{M_{\rm{Jup}}})$  & 0.467 & 0.041 & B06 \\
      $a~ (\rm{AU})$ & 0.0482 & 0.0028 & B06 \\
      $e$ & 0.034 & 0.029 & B06 \\
      $K_{\rm{s}}~(\rm{km~s^{-1}})$ & 0.0549 & 0.0019 & B06 \\
      Orbital period (d) & 3.509267 & 0.000064  & B06 \\
      ${ T_{\phi=0}}$ (JD) & 2~450~829.872 & 0.038  & B06 \\
      \hline                                   
    \end{tabular}
    \end{center}
    References: B06 = Butler et al. (2006) and references therein,
    G89~=~Gratton et al.~(1989), U00 = Udry et al.~(2000), VF05~=~Valenti \& Fischer (2005).
  \end{table}

  The planetary system of HD~75289A seemed to bear the opportunity to detect the reflected
  starlight for the following reasons: First, the brightness of HD~75289A in
  the visual ($V=6.35$ mag) enables a large amount of high-S/N spectra to be
  acquired within a short
  period of time, which helps to reduce photon noise, the dominant noise source. 
  Second, the amount of starlight
  received from this planet is likely to be high due to the extremely
  small distance to its host star of only $a=0.048$~AU (cf. Equation \ref{equ:2}).
  Third, a G0 V star shows 
  a rich absorption-line spectrum at optical wavelengths, a prerequisite for
  the data-synthesis approach outlined in Section 4. 

  Using data of the stellar mass $M_{\rm{s}}$, the planetary minimum
  mass $M_{\rm{p}}\sin i$, and the
  RV semi-amplitude of the reflex motion of the star $K_{\rm{s}}$, 
  the maximum possible RV semi-amplitude of the planet can be determined (from
  Equation~\ref{equ:doppler}) to be $K_{\rm p,max}=148.9\pm15.8~\rm{km~s^{-1}}$. 
  An estimate of the planetary radius $R_{\rm{p}}$ can be obtained from a
  comparison of the radii determined for other hot Jupiters with the transit
  method. To this end, we selected a subsample of known transiting hot
  Jupiters  (Burrows et al. 2007, and references therein), based on the following criteria:
  (i)~Planet mass between the minimum mass of HD~75289Ab and 2.3 times
  this value (for random orientation of the planetary orbit there is a 90~\%
  confidence that the true mass of HD~75289Ab lies in this interval).
  (ii)~Orbital radii similar to that of HD~75289Ab within 20~\%.
  (iii)~Host star spectral types close to G0 V (F8 V~-~G1 V). The resulting
  group of objects
  contains OGLE~10b, OGLE~111b, HAT-P-1b, XO-1b, HD~209458b. From this sample, we estimated the
  planetary radius of  HD~75289Ab as the average of the radii of these five hot
  Jupiters finding $R_{\rm{p}}=1.2 \pm 0.14 ~R_{\rm Jup}$.  The uncertainty
  corresponds to the scatter in the above sample of transit radii.


   \section{Data}
   \begin{table*}
     \caption{Journal of observations. The UTC times and the orbital phases of
   the planets are shown. Note that  $\phi_{\rm{Leigh}}$ indicates the erroneous phases
   applied by Leigh et al. (2003), while  $\phi_{\rm{corr}}$ refers to the corrected
   ones. The number of spectra obtained is given in the last
   column. Ephemerides used are: orbital period = 3.509267~d, 
   ${T_{\phi=0}}$ (BJD) = 2~450~829.872 (Butler et al. 2006, and references therein).}             

 \centering                          

     \begin{tabular}{c c c c c c c c }        
       \hline\hline                 
       Night & UTC start &  $\phi_{\rm{Leigh}}$ &  $\phi_{\rm{corr}}$ & UTC end &  $\phi_{\rm{Leigh}}$ &  $\phi_{\rm{corr}}$ & $N_{\rm{spectra}}$  \\
       \hline                        
        1 & 2003/01/14~~01:00 & 0.39 & 0.68 & 2003/01/14~~09:30 & 0.50 & 0.79 &
        188 \\
        2 & 2003/01/15~~00:57 & 0.68 & 0.97 & 2003/01/15~~09:22 & 0.78 & 0.07 &
        173 \\
        3 & 2003/01/21~~02:40 & 0.41 & 0.70 & 2003/01/21~~09:35 & 0.49 & 0.78 &
        183 \\
        4 & 2003/01/22~~03:36 & 0.71 & 0.00 & 2003/01/22~~09:41 & 0.78 & 0.07 &
        140 \\
       \hline                                   
     \end{tabular}
     \label{tab:pha}      
  \end{table*}

   We retrieved 684 high-resolution and high-S/N spectra 
   files from the ESO science archive, along with the calibration
   files. The observations were conducted over four nights by Leigh
   et al. (2003a) with UVES mounted
   on the VLT/UT2. The blue arm of the
   spectrograph was used with the single EEV CCD-44
   array detector centered at wavelength $\lambda = 475.8~\rm{nm}$ providing 25 full
   orders and covering the wavelength range $\lambda=402.3$ to $522.1
   ~\rm{nm}$. The integration times were adapted to the seeing such
   that a count rate
   not exceeding 40~000~ADU per 
   pixel was achieved, insuring high S/N levels and at the same time staying
   clear of the saturation level. The effective resolving power was $R=43~000$. More details of the
   observations can be found in Leigh et al. (2003a).

   These authors selected the observing dates for the best visibility of the planet,
   which is in the orbital phase range  $\phi=0.30$ to $0.45$ and $0.55$ to
   $0.70$ (close to superior
   conjunction), but excluding phases about 0.50, where the absorption-line
   systems of the star and the planet blend. 
   However, we found that a one-day error in the date used for the phase
   calculation must have been made. This resulted in the fact that the planet
   was not observed at the optimum phases. In particular, 
   during two nights the observations were 
   taken close to the inferior conjunction of the planet,
   where it is hardly visible. Furthermore, the same
   erroneous phases were used in the subsequent data analysis. 
   Consequently, one can expect the corrected upper limit to the
   planet-to-star flux ratio to be smaller than the value determined by Leigh et
   al. 
   Table \ref{tab:pha} shows the journal
   of observations with the corresponding phases of the planet (erroneous and
   corrected ones).  


   \subsection{Data reduction}

 The data were reduced using a UVES data-reduction pipeline developed by
   ourselves. For high-quality
   flat-field correction, 228 and 369 flat field
   exposures were combined for the first two nights and for the last
   two ones, respectively. Furthermore, 90 bias exposures were combined.
   We retrieved 25 orders
   of 3000 pixels from each echelle spectrum. No order merging was applied.
   Furthermore, we identified cosmic-ray hits by way
   of the following procedure: for each spectrum, we
   compared the flux in every pixel with the median flux of the same pixel in
   the three predecessor and the
   three successor spectra, which had been scaled to the same flux as the
   spectrum under consideration. We flagged those pixels where the 
   difference exceeded $6 \sigma$ as cosmic-ray hits. These pixels were 
   then excluded from further analysis.
   
   We discarded the most weakly exposed regions of each echelle order 
   (the first 300 pixels as well as the last 100
   pixels).
   To speed up the data analysis, we co-added spectra into
   sufficiently narrow phase bins
   such that phase smearing of the stellar absorption lines 
originating from stellar RV
   variation, sub-pixel shifts (see Section \ref{S4.1}), and the
   barycentric motion of the Earth, remained below 0.04~${\rm km~s^{-1}}$. An
   additional criterion for the size of the phase bins was that the
   unseen planetary lines did not suffer from smearing in excess of $2~{\rm km~s^{-1}}$.
 In these ways, we reduced the number of spectra from 684 to 210.

   \section{Data analysis: The data synthesis method}

   We model the starlight reflected from the planet as a copy of the stellar spectrum,
   strongly scaled down in brightness and Doppler-shifted according to the orbital motion of the
   planet. The HD~75289A spectra have an average S/N of $300$ to $600$ per
   dispersion element. With the expectation the planet-to-star flux ratios
   of the order of a few times $10^{-5}$, it is clear that the reflected spectrum from the planet is deeply buried in the noise of the stellar spectrum. The weak planetary
   signal is increased by the large number of spectra, and more importantly, 
   by the combination of the approximately $1500$ absorption lines, achieved using the  data-synthesis method described below. 

   \subsection{Step 1: Construction of the superspectrum} \label{S4.1}
   To create a  high S/N, virtually planet-free superspectrum,
   we first create copies of the original, unmodified object spectra. These
   copies are then co-aligned with respect to the barycentric velocity of the Earth
   and the radial velocity of the star, and finally summed up. Here, it is
   important that the observing dates cover different orbital phases well
   distributed such that the planetary signal in the
   superspectrum is washed out. In the case of the HD~75289A data, nearly half of
   the observations were
   conducted in the orbital phase range
   $\phi=0.97$ to $0.07$ (Table \ref{tab:pha}),
   where the planet was close to
   inferior conjunction and therefore at its faintest (e.g. for an orbital
   inclination of $i=60^{\circ}$, the planet appears about 12 times fainter at the
   inferior conjunction, relative to its maximum brightness).
   Consequently, we used only these ``planet-free'' spectra to create the superspectrum.
   
 We found that imperfections in the dispersion solution (sub-pixel
 shifts and stretches/contractions) of the different 
 spectra led to broadening of absorption lines in the superspectrum. 
 We assume that these shifts originate mostly from errors of the
 wavelength calibration and probably to some extent from guiding errors
 and variations in the instrumental profile. 
 Consequently, we correct each spectrum for these wavelength errors
 before producing a new version of the superspectrum.  To this end, we first divide
 each order of the modified copies of the object spectra into chunks of size $N$ pixel, and
 determine the central wavelength $\lambda_{i}$ of each chunk $i$. For the data analysis of
   HD~75289A, we set
   $N=120$, which creates 22 chunks per order. By means of
   the Brent algorithm (Press et al. 1992), we determine for each
   chunk the sub-pixel shift $s_{i}$ between the superspectrum and
   the modified copy of the object spectrum under consideration.  Then, we
 shift each pixel of 
 the modified copy by a value determined by calculating
 a spline through 
 all pairs of $\lambda_{i}$, $s_{i}$ per spectral order.  Finally, we add up the re-adjusted
  co-aligned versions of the copies of the object spectra and get an improved version of the
 superspectrum.

    \subsection{Step 2a: Modeling the stellar signal ... }

     Each of the original, unmodified object spectra is modeled using the improved
    superspectrum. In this initial step, we ignore the presence of the faint
    planetary signal
    in the data. First the superspectrum (model) is corrected for a general
    linear trend in flux. Second, the model is shifted according to the
    barycentric velocity of the Earth, the radial velocity of the star, and the
    aforementioned shifts and stretches/contractions in the
    sub-pixel regime, so that the
    positions of absorption lines of the object spectrum and the model are
    matched. This is achieved using a chunk-Brent-spline-approach similar to
    that described in
    the previous paragraph. We note that in all analysis steps, modifications are
    exclusively applied to the model, but the object spectrum are used
    in their original version, i.e. the data are unchanged.

    At the third stage, the model is scaled chunkwise with respect to the object
    spectrum. 
 As we now compare the scaled model with each object spectrum, we notice that the
    widths and depths of the absorption lines differ slightly. These
    differences originate most likely in the aforementioned effects of
    residuals wavelength calibration errors, guiding errors, and variations in the
 instrumental profile, and can
    be corrected by adding a scaled version of the second derivative of the object
    spectrum to the model. The scaling factor is determined via $\chi^2$
    minimisation (fourth stage). After this, the model is
    renormalised. The final stage is to 
    iterate twice over all these four  
    processes to improve the model describing the stellar spectrum.
 
     \subsection{Step 2b: ...  and the planetary signal}
     
     For the model of the planetary signal, we use a copy of the improved model of the
     stellar spectrum, but scaled down by the factors  $\epsilon(\lambda)~\mu(\phi, i)$ and
     shifted by velocity $V_{\rm{p}}(K_{\rm{p}},\phi)$ with respect to the
     stellar spectrum. Hence, the two free parameters are the planet-to-star
     flux ratio for the fully-illuminated planet $\epsilon(\lambda)$, and the
     orbital inclination $i$, which corresponds to the 
     RV semi-amplitude of the planet $K_{\rm{p}}=K_{\rm{p,max}} \sin i=148.9
     \sin i~~{\rm km~s^{-1}}$.

   \subsection{Step 3: Evaluation of the models}

   We are now ready to add this planetary signal to the improved model $T$ of the
   stellar spectrum and consequently construct the model $M$ describing
     the spectrum of the star {\it plus} the reflected one from the
     planet. For each pixel $k$, $M$ is given by
 \begin{equation} \label{E4.2a}
    M_{k} = \frac{T_{k}(\lambda_k)+\epsilon(\lambda_k)~\mu(\phi,
    i)~T_{k}\lbrace\lambda_k~[1+V_p(K_{\rm p},\phi)~c^{-1}]\rbrace }{1+\epsilon(\lambda_k)~\mu(\phi, i)} {\rm,}
 \end{equation}
  where $c$ denotes the speed of light. Varying $K_{\rm p}$ and $\epsilon(\lambda)$,
  we finally search for the best-fit model $M$ to all the object spectra by $\chi^2$ minimisation. The search range for the RV
  semi-amplitude comprised $K_{\rm p} = 40$ to $180~{\rm km~s^{-1}}$
  (corresponding to orbital inclinations $i=15^\circ$ to $90^\circ$, plus
  twice the error of $K_{\rm p,max}$; see Table~1) with a step
  width of $3~{\rm km~s^{-1}}$. This was a good compromise between computing
  time and sampling the  average absorption line profile with the FWHM of $\approx
  15~{\rm km~s^{-1}}$. Using simulations, we found that for small
  inclinations of the planetary orbit, where the planets appear only slightly
  illuminated, the method
  is unable to detect Jupiter-size objects with very high albedos.

   \subsection{Step 4: Determination of the confidence level}
   Once the best model $M[K_{\rm p},\epsilon(\lambda)]$ has been evaluated, we 
   determine the confidence level of the $\chi^2$ minimum by applying the
   bootstrap randomisation method (e.g. K\"urster et al.~1997). Retaining the
   orbital phases, we randomly redistribute the observed
   spectra amongst the phases, thereby creating $N$ different data sets.
   Any signal present in the original data is now scrambled in these
   artificial data sets.
   For all these randomised data sets, we again evaluate the model for the
   two free parameters, and locate the best fit with its specific $\chi^2$
   minimum. We set $m$ to be the number of best-fit models to the $N$ randomised
   data sets that have a minimum $\chi^2$ less or equal than the minimum $\chi^2$ found for the original
   data set. The confidence level can then be estimated by $\approx 1-m/N$.

   \section{Results and discussion} 
   
   Applying the data synthesis method to the HD~75289A data, we adopted
   the following approximations to the atmospheric models by Sudarsky et al.~(2000). 
   \vspace{2mm}

   (i)  We adopted a grey-albedo model to resemble the Class V
   model, which describes the atmospheres of hot Jupiters with temperatures $>1500~{\rm
   K}$. 
As can be seen in Figure~\ref{fig:albedo}, this was a valid approximation for
   our observed wavelength range
   (402 to 522 nm). 

(ii) We also considered the two Class IV 
   models for hot Jupiters
   with temperatures around $1300~{\rm K}$. The irradiated model predicts a very low reflectivity
   due to the lack of silicates in the upper layers of the planetary
   atmosphere, while the isolated model predicts high reflectivity due to
   the existence of silicate clouds in the upper decks of the planetary
   atmosphere. We found that both Class IV models show a similar trend for the albedo
   function.   Therefore we 
   considered only that trend in further modeling.

   \subsection{Grey albedo model (corresponding to Class V)} \label{E:grey}
\begin{figure}
  \centering \includegraphics[bb=30 50 550
  780,angle=-90,width=9cm]{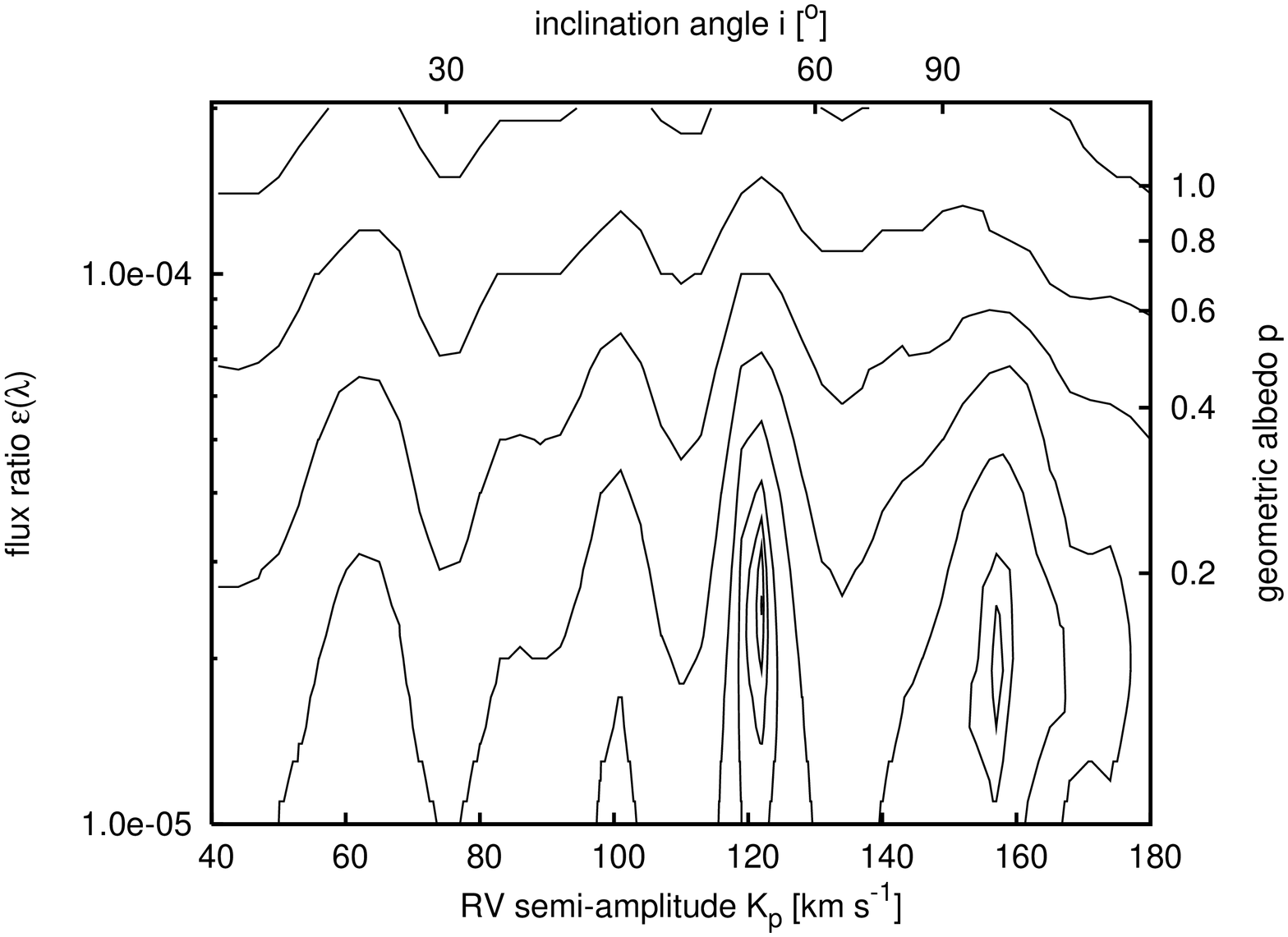}
  \includegraphics[bb=30 50 550 780,angle=-90,width=9cm]{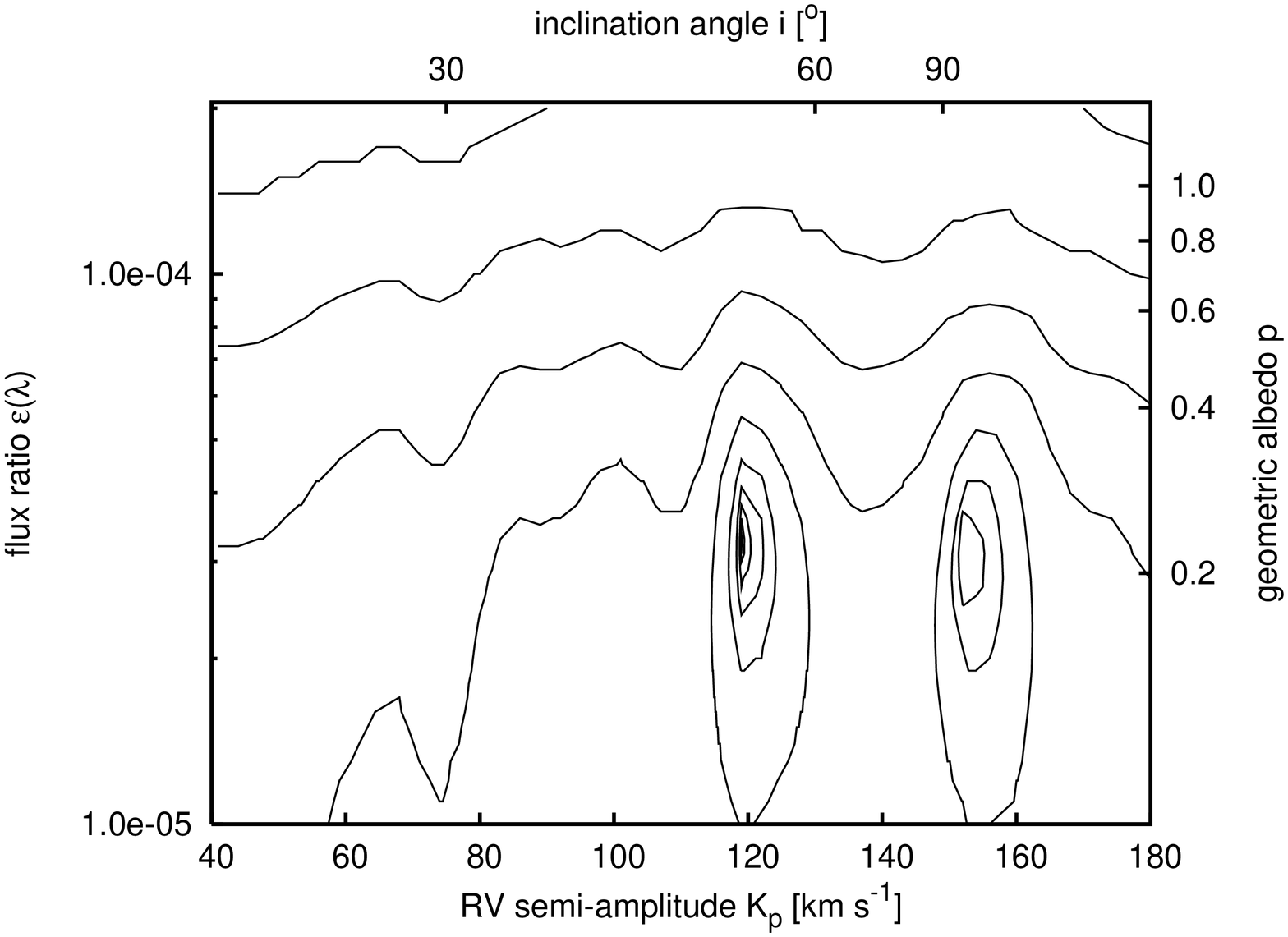}
      \caption{Contour maps of $\chi^2$ normalised to $\chi^2_{\rm min}$ for 
        the model parameters $K_{\rm{p}}$ and
   $\epsilon(\lambda)$ showing the result of the analysis of the HD~75289A
   data for two different atmospheric models.
   For better visualisation the $\chi^2$ contour levels follow the
  sequence $\chi^2_k = \chi^2_{\rm min}+ 0.1~\chi^{-2}_{\rm min}~e^{k-1}$ for $k>0$. \newline
   Upper panel:
   For the planetary model, we assumed a Venus-like
  phase function and a grey albedo. The minimum $\chi^2$ was found at
  $K_{\rm{p}}=121~\rm{km~s^{-1}}$ and
  $\epsilon(\lambda)=2.9\times10^{-5}$. 
 An analysis with 3000 trials revealed that this feature is uncertain with
  a false alarm probability of $19\%$. \newline
   Lower panel:
 For the planetary model, we assumed a Venus-like
  phase function and a class IV albedo function. 
  The minimum $\chi^2$ was found at
  $K_{\rm{p}}=119~\rm{km~s^{-1}}$ and
  $\epsilon(\lambda)=3.3\times10^{-5}$. 
  An analysis with 3000 trials revealed that this feature is uncertain with
  a false alarm probability of $24 \%$. 
 We note that the corresponding geometric albedo $p$ is shown under the
   assumption that $R_{\rm p} = 1.2~R_{\rm Jup}$ (the right-hand
   y-axis).
 }
         \label{fig:nodetection}
   \end{figure}


   Our data analysis, using a grey albedo, revealed  
   a $\chi^2$-minimum at a planet-to-star flux ratio $\epsilon(\lambda) =
   2.9\times10^{-5}$, and an RV semi-amplitude
   $K_{\rm{p}}=121~\rm{km~s^{-1}}$ corresponding to an orbital inclination of
   $i=54^{\circ}$. Figure \ref{fig:nodetection} (upper panel) shows a $\chi^2$
   contour map.  However, using bootstrap randomisation with 3000
   trial data sets we found that this $\chi^2$-minimum was uncertain with a
   false-alarm probability (FAP) of 19 \%; we do not therefore consider this as a
   detection of reflected light from the planet.


   \subsection{Class IV model}


   The analysis using the Class IV albedo model 
   did not yield  any evidence for reflected light from the hot Jupiter
   HD~75289Ab. We found that the $\chi^2$-minimum at a planet-to-star flux 
   ratio $\epsilon(\lambda) = 3.3 \times
   10^{-5}$ and an RV semi-amplitude 
   $K_{\rm{p}}=119~\rm{km~s^{-1}}$, which corresponds to an orbital
   inclination $i=53^\circ$. Figure \ref{fig:nodetection} (lower panel) shows a 
 contour map of that feature. From bootstrap randomisation with 3000
   trials we found that this feature was uncertain with a FAP of 24\%.

\subsection{Upper limits}
 
  \begin{figure}
   \centering \includegraphics[bb=30 50 550 780,angle=-90,width=9cm]{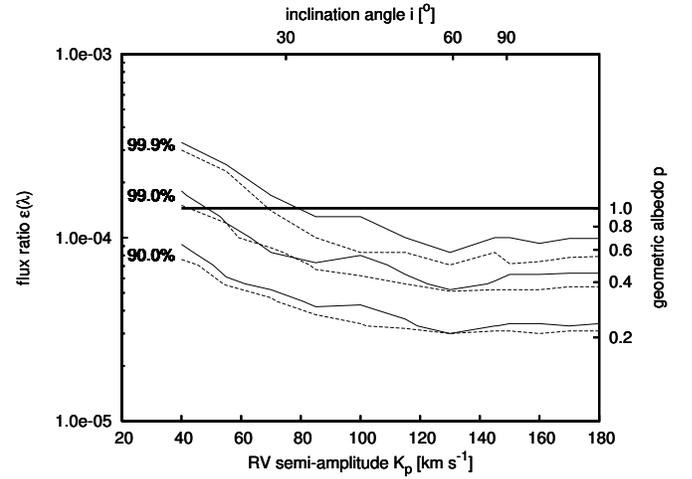}
      \caption{Contour map showing confidence levels for the upper limits to
   the planet-to-star flux ratio $\epsilon(\lambda)$ as a function of to the 
   RV semi-amplitude
   $K_{\rm{p}}$ of the planet (lower x-axis), or orbital inclination $i$
   (upper x-axis). 
   From top to bottom, these levels are:
   99.9, 99.0 and 90.0 \% confidence. For the planetary model, we assumed a Venus-like
   phase function as well as a Class IV albedo function (solid lines) and a grey
   albedo (dashed lines). The corresponding geometric albedo $p$ is shown under the
   assumption that $R_{\rm p} = 1.2~R_{\rm Jup}$ (the right-hand
   y-axis). The region above the thick horizontal line is excluded because the
   albedo of the planetary atmosphere would exceed 1.
}
         \label{fig:upperlimit1}
   \end{figure}

  Obviously, the amount of data for when the planet was observed during its bright phases 
   was insufficient to measure the  planetary signal. The question is now:
   at what planet-to-star flux ratio $\epsilon(\lambda)$ would we be able to
   detect significantly the planetary signal using the present data?
   To this end, we constructed data sets by adding an artificial
   planetary signal to the original data. To produce the planetary signal, we explored the range of
   $K_{\rm{p}}\in[40,~180]~{\rm km~s^{-1}}$ and
   $\epsilon(\lambda)\in[2\times10^{-5},~2.5\times10^{-4}]$. For each data set, 
   the simulated planetary signal was a copy
   of the object spectrum, but shifted by
   $V_p(K_{\rm{p}},\phi)$ and scaled down by the factor $f(\phi, i,
   \lambda)$, which includes the albedo model of the planetary atmosphere
   (either grey albedo or Class IV model). For each pair of the
   parameters $K_{\rm{p}}$ and $\epsilon(\lambda)$, we determined the
   confidence level of the best-fit model to the artificial data
   via bootstrap randomisation.

   This enables place upper limits to be established to the planet-to-star flux
   ratio $\epsilon(\lambda)$ for different confidence levels, as a function of 
   the RV semi-amplitude $K_{\rm p}$. Figure~\ref{fig:upperlimit1} shows that
   the upper limits to the planet-to-star flux
   ratio decrease with increasing orbital inclination, which is a direct
   consequence of the illumination geometry. 
   As can be seen from Figure~\ref{fig:upperlimit1} (dashed lines), we have
   the highest sensitivity for detection of the planetary signal at high
   orbital inclinations with the formal numerical optimum found at an 
   RV semi-amplitude $K_{\rm p}=129~{\rm km~s^{-1}}$ corresponding to
   $i=60^\circ$.  At this inclination, the 99.9~\% confidence
   upper limit to the planet-to-star flux ratio for the grey albedo model is
   $\epsilon=6.7\times10^{-5}$ , while it is  $\epsilon=8.3\times10^{-5}$  for
   the Class IV model. In Figure~\ref{fig:upperlimit1}, we clearly see
   that the upper limits established by adopting the grey albedo model are
   deeper than the ones found with the Class IV model. A plausible explanation
   for this seems to be that in the grey albedo model, all the absorption lines
   have the same weight.  On the other hand, in the Class IV model, the
   absorption lines in the red part of the observed spectrum have a lower
   weight than those in the blue part. Hence, fewer lines are effectively used for
   the analysis.

 Assuming a planetary radius $R_{\rm p}=1.2~R_{\rm Jup}$, we find that the 99.9 \%
   upper limit to the geometric albedo is $p<0.46$ for the grey albedo
   model, and  $p<0.57$ for the Class IV model, at a wavelength 
   of $\lambda=445~{\rm nm}$, which  corresponds to the center of gravity of the
   albedo function. For comparison, at this wavelength Jupiter's albedo is $p=0.43$
   (Karkoschka 1994). We note that the planetary radius of $R_{\rm p}=1.2~R_{\rm
   Jup}$ is the average value of the five transiting hot Jupiters listed in
   Section 2.3. 

   In the case that the planetary radius of HD~75289Ab is similar to that
   of the transiting hot Jupiter HAT-P-1~b 
  ($R_{\rm p}=1.36~R_{\rm Jup}$, Bakos et al. 2007), we find the
  geoemetric albedo to be $p<0.36$ and $p<0.45$, respectively for the grey albedo model and
   the Class IV model. This would enable us to rule out the Class V
   albedo model.

   \begin{table}
     \caption{Estimates for the observing times required for UVES to collect high-S/N data
       sufficient to detect, with 99.9\%
       confidence, the reflected light from HD~75289Ab with a
       specific planet-to-star flux ratio $\epsilon$.   Here we assume a Class
       V model for high-reflective case and the isolated Class IV albedo shape for
       the fainter case (cf. Fig. 1). 
       The corresponding values of the albedo values $p$ are based on the assumption that
       the planetary radius is $R_{\rm p}=1.2~R_{\rm Jup}$. 
       }

 \centering                          

     \begin{tabular}{c c c c l}        
       \hline\hline                 
       $\epsilon$ & p & N (spectra) & time (h) & albedo model\\
       \hline
       $5.7\times10^{-5}$ & 0.4 & 324 & 19.5 & Class V \\
       $2.9\times10^{-5}$ & 0.2 & 1460 & 87.6  & isolated Class IV\\
       \hline                                   
     \end{tabular}
     \label{T:3}      
  \end{table}


   \section{Conclusions and outlook}

   We reanalysed 684 high-precision spectra of the star HD~75289A, obtained
   during four nights by
   another research group in 2003, in an attempt
   to detect starlight reflected from the known hot Jupiter. 
   \vspace{2mm}

   (i) We found that the upper limits to the planet-to-star ratios established by
   Leigh et al.~(2003a) were based on erroneous orbital phase information, and therefore
   needed to be corrected. 

   (ii) Our reanalysis using the correct orbital phases produced a non-detection of the reflected light of the
   planet. We therefore placed upper limits on the planet-to-star flux ratios for
   different possible orbital inclinations and confidence levels. These upper
   limits, however, were $\approx 60$~\% higher than the ones placed by Leigh
   since the planet was not 
   observed at its best visibility. 

   (iii) We furthermore determined values of the geometric albedo that was 2.9
   times higher than the values established by
   Leigh et al. (2003a), since, due to limited information on planetary
   transits available in 2003, these authors assumed a planetary
   radius that was larger by 33~\% than our value. 

   (iv)   The previous upper limit to the geometrical albedo of 0.12
   established by Leigh et al. (2003) was by far the deepest one determined
   for hot Jupiters up to now. These authors suggested that a Class V atmospheric model
   can be ruled out for HD~75289Ab. For the value of the planetary radius
   adopted by us, $R_{\rm p}=1.2~R_{\rm Jup}$, we find that the amount of data was
   insufficient to rule out specific atmospheric models. However, if the
   radius of HD~75289Ab is similar to that of the transiting planet HAT-P-1~b 
  ($R_{\rm p}=1.36~R_{\rm Jup}$, Bakos et al. 2007), we can rule
   out the Class V albedo model.
   \vspace{2mm}

    To detect reflected starlight from the hot Jupiter HD~75289Ab, 
    more high-precision data would be needed.  Table~\ref{T:3} lists  estimates of
    observing times required to detect, with 99.9\% confidence, the reflected
    light of a planet with a Class V model and an isolated Class IV model, respectively.
    These calculations are based
    on the following assumptions: the fictitious observations are scheduled
    optimally with respect to phase ($\phi=0.30$ to $0.45$ and $0.55$ to 
   $0.70$). These observations are carried out by using the blue arm of UVES
    at the VLT/UT2 in Chile. The wavelength range is
    $\lambda=400.3$ to $527.0~\rm{nm}$, and the exposure time is chosen in
    such a way that a maximum count rate of 40 000 ADU/px is achieved
    (cf. Section 3). Table~\ref{T:3} demonstrates that we would require about 20 hours of
    observations to
    measure the reflected light of HD~75289Ab in the case that it had a Class V albedo
    function and a planetary
    radius of $R_{\rm p}=1.2~R_{\rm Jup}$. However, if HD~75289Ab had an
albedo function similar to that of the isolated Class IV model, it would
    already take more than 87 hours of
    observing time to detect the reflected light.

    Because of the large amount of data required, we conclude that the star
    HD~75289A is too faint for the search for reflected starlight. 
    Instead, the search for thermal emission of HD~75289Ab in the
    near-infrared appears
    to be promising, since the
    planet-to-star flux ratio increases dramatically in the infrared (Barnes et al. 2007).

%


   \begin{acknowledgements}

   \end{acknowledgements}
   We are grateful to Tsevi Mazeh 
   for valuable discussions. Furthermore, we want to thank the anonymous
   referee for valuable comments. Based on observations made with ESO Telescopes 
   at the Paranal Observatory under programme ID 70.C-0061(A).



\begin{thebibliography}{}
   
\bibitem[2007]{bakos07} Bakos G., Noyes R., Kov\'acz G., Latham D., et
  al. 2007, ApJ Letters, 656, 552

\bibitem[2007]{barnes07}Barnes J. R., Leigh C. J., Jones H. R. A., et
  al. 2007, MNRAS, 379, 1097

  \bibitem[2007]{burrows07} Burrows A., Hubeny I., Budaj J., Hubbard W. B., 2007, ApJ, 661, 502

     \bibitem[2006]{butler06} Butler R. P., Wright J. T., Marcy G. W., Fischer
     D. A., Vogt S. S., Tinney, C. G., 2006, ApJ, 646, 505

   \bibitem[1999]{charb} Charbonneau D., Noyes R.W., Korzennik S.G., Nisenson
   P., Jha S., Vogt S.S., \&~Kibrik R.I., 1999, ApJ, 522, L145


 \bibitem[1989]{gra89} Gratton R.~G., Focardi P., Bandiera R., 1989, MNRAS, 237, 1085

   \bibitem[1992]{hilton92} Hilton, J.L., 1992, Explanatory Supplement to the
     Astronomical Almanac, Universiy Science Books, Mill Valley CA, p. 383

   \bibitem[1994]{Karkoschka} Karkoschka E., 1994, Icarus, 111, 174
   \bibitem[1997]{Kur97} K\"urster M., Schmitt J.H.M.M., Cutispoto G., Dennerl
   K., 1997, A\&A, 320, 831

   \bibitem[2003a]{leigh} Leigh C., Collier Cameron A., Horne K., Penny A., 2003a, MNRAS, 344, 1271


   \bibitem[2003b]{leigh} Leigh C., Collier Cameron A., Guillot T., 2003b, MNRAS, 346, 890




   \bibitem[2004]{Mug04} Mugrauer M., Neuh\"auser R., Mazeh T., Alves J.,
    2004, A\&A, 425, 249


 \bibitem[1997]{Kur97} Press W.~H., Teukolsky S.~A., Vetterling W.~T.,
 Flannery B.~P., 1992,   Numerical recipes in C. The art of scientific
 computing, Cambridge University Press
   
   \bibitem[1998]{sea98} Seager S. \& Sasselov D. D., 1998, ApJ, 502, 157

   \bibitem[2000]{Sudarsky00} Sudarsky D., Burrows, A., Pinto P., 2000, ApJ,
   538, 885
        
  \bibitem[2000]{Sudarsky03} Sudarsky D., Burrows A., Hubeny I., 2003, ApJ,
  588, 1121

  \bibitem[2000]{Udry00} Udry S., Mayor M., Naef D., Pepe F., Queloz D., 2000, A\&A, 356, 590 

   \bibitem[2005]{VF05} Valenti J. A., Fischer D. A., 2005, ApJ Supl., 159, 141
\end{thebibliography}
\end{document}